# Verification of the flow regimes based on high fidelity observations of bright meteors

Manuel Moreno-Ibáñez[1], Elizabeth A. Silber[2], Maria Gritsevich[3,4], Josep M. Trigo-Rodríguez[1,5]

[1]*Institute of Space Sciences (ICE, CSIC), Meteorites, Minor Bodies and Planetary Science Group, Campus UAB, Carrer de Can Magrans, s/n E-08193 Cerdanyola del Vallés, Barcelona, Catalonia, Spain*
mmoreno@ice.csic.es

[2]*Department of Earth, Environmental and Planetary Sciences, Brown University, Providence, RI, 02912, USA*
elizabeth_silber@brown.edu

[3]*Department of Physics, University of Helsinki, Gustaf Hällströmin katu 2a, P.O. Box 64, FI-00014 Helsinki, Finland.*

[4]*Institute of Physics and Technology, Ural Federal University, Mira str. 19. 620002 Ekaterinburg, Russia.*

[5]*Institut d'Estudis Espacials de Catalunya (IEEC), C/ Gran Capitá, 2-4, Ed. Nexus, desp. 201, E-08034 Barcelona, Catalonia, Spain*

**Abstract**

Infrasound monitoring has proved to be effective in detection of the meteor generated shock waves. When combined with optical observations of meteors, this technique is also reliable for detecting centimeter-sized meteoroids that usually ablate at high altitudes, thus offering relevant clues that open the exploration of the meteoroid flight regimes. Since a shock wave is formed as a result of a passage of the meteoroid through the atmosphere, the knowledge of the physical parameters of the surrounding gas around the meteoroid surface can be used to determine the meteor flow regime. This study analyses the flow regimes of a data set of twenty-four centimeter-sized meteoroids for which well constrained infrasound and photometric information is available. This is the first time that the flow regimes for meteoroids in this size range are validated from observations. From our approach, the Knudsen and Reynolds numbers are calculated, and two different flow regime evaluation approaches are compared in order to validate the theoretical formulation. The results demonstrate that a combination of fluid dynamic dimensionless parameters is needed to allow a better inclusion of the local physical processes of the phenomena.

# 1. INTRODUCTION

Studies of meteoroids entering the Earth's atmosphere offer insight into the characteristics of these objects, as well as the conditions under which they produce shock waves. Despite recent advancements in meteor science, the classically derived flow regimes of meteoroids in the centimeter size range have never been validated against a well constrained observational data set. Validation and better characterization of the flow regimes associated with bright meteors are essential for considerations of the onset of shock waves produced by these objects in the upper atmosphere, as well as for developing new atmospheric flight models, the examination of ablation processes assumptions and the improvements of the studies derived from meteor observations. Furthermore, these may have implications on other scientific areas such as: aeronomy, shock physics, meteor and near-Earth Object (NEO) research, and planetary defence studies.

## 1.1 Flow regimes

Meteoroids are solid objects which originate from comets, asteroids and other solar system bodies. Their orbits are perturbed by the gravitational influences of planets, or due to collisions (Jenniskens 1998; Trigo-Rodríguez et al. 2005a; Trigo-Rodríguez et al. 2005b; Dmitriev et al. 2015). Meteoroids impact the Earth's atmosphere at hypersonic entry velocities, ranging between 11 - 73 km/s, corresponding to the Mach number ($Ma$), which represents the ratio of the meteoroid velocity to the local speed of sound at the meteoroid surrounding flow conditions, between 35 and 270 (e.g., Ceplecha et al., 1998; Jenniskens 1998; Baggaley 2002; Gritsevich 2009). If large and capable of depositing sufficient energy, these objects can generate shock waves that in some cases might produce destructive effects on the ground (e.g., Brown et al. 2013b; Tapia & Trigo-Rodriguez 2017).

Upon encountering the Earth's atmosphere, the meteoroid generates light (due to friction with air molecules followed by ionization, ablation, sputtering, and fragmentation), eventually producing a bright column of ionized gas called a meteor.

On its passage through the atmosphere, the meteoroid encounters increasing gas density and thus an increasing number of impinging particles. However, the number and energy of the impinging particles are not only a function of the gas density at the corresponding height, but are also related to the velocity and the size of the body. The kinetic energy of the impinging particles depends on the Mach number. This results in several possible physical flight scenarios known as the flow regimes. There are four commonly accepted flow regimes: free-flow, transitional, slip-flow and continuum. These are characterized by a dimensionless parameter called the Knudsen number ($Kn$), which is defined as the ratio between the mean free path of the gas molecules ($\lambda$) to a characteristic length scale ($L$) of the body immersed in the gas, and thus $Kn=\lambda/L$. It is quite common to use an equivalent radius of the meteoroid ($r$) as the characteristic length (e.g. Gritsevich & Stulov 2006). However, when a boundary layer exists (a region in the vicinity of the body where the viscous effects are significant), the thickness of the boundary layer ($\delta$) is used as the characteristic scale, $Kn=\lambda/\delta$ (Bronshten 1965, 1983). Alternatively, the $Kn$ number can be described as the inverse product of the intermolecular collision rate ($v$) and a characteristic flow time ($t$), thus $Kn=1/(v \cdot t)$. The latter definition demonstrates that the larger the number of the collisions for a given time, the smaller the $Kn$ value. Note that the collision rate applies only to the gas molecules; the collisions against the body surface are not accounted for in this scenario. The rate of collisions controls the distribution of velocities of the impinging molecules and thus the mathematical formulation to be applied to the physical scenario. This eventually hinders a sharp delineation of the flow regime limits, since it is not trivial to constrain the molecular collision rate at each stage of the meteoroid's descend through the atmosphere.

The first $Kn$ expression, $Kn = \lambda/L$, is the most common and practical, although defining $\lambda$ can be challenging as its definition is not unique, and it can be regarded differently owing to the molecules and the reference frame considered in a given study. As explained in Bronshten (1983), there are more than eight possible scenarios, out of which, two are usually the most commonly adopted. On the one hand, blunt bodies (i.e., re-entry vehicles) are generally studied using a reference frame moving with the gas and the equilibrium air molecules. On the other hand, as discussed by Rajchl (1969) and Bronshten (1983), for meteor problems where the immersed body loses material during its movement and the shape of the meteoroid is not known, it is more realistic to fix the reference frame to the meteoroid and study the mean free path of the reflected (or evaporated) molecules relative to the impinging molecules. Furthermore, this approach allows a separate analysis of the various local scenarios in the vicinity of the meteoroid (Josyula & Burt 2011). To make a distinction between these scenarios, the latter $Kn$ is renamed to $B$ (Rajchl 1969) or $Kn_r$ (Bronshten 1983). Hereafter, the nomenclature $Kn_r$ will be adopted to refer to this second definition of *Kn* approach, where the reference frame is fixed to the meteoroid.

There are various flow regime classifications based only on *Kn* or a combination of *Kn* with other parameters. The most widely used classification (hereafter referred to as the classical scale) accounts for the number of intermolecular collisions in a specific time (recall that *Kn* is proportional to the inverse product of the intermolecular collision rate); it is as follows:

i) Free molecular regime, $Kn>10$. The number of intermolecular collisions is scarce. Single molecules hit the immersed body;

ii) Transitional regime, $0.1<Kn<10$. The mean free path of the molecules is of the same order of magnitude as the body characteristic size. There are collisions between molecules;

iii) Slip-flow regime, $0.01<Kn<0.1$. There is a slightly tangential component of the flow velocity in the boundaries of the body's surface, but there is no adhesion of the flow to the body's surface;

iv) Continuum-flow regime, $Kn<0.01$. The flow is considered to be continuous.

Another typical strategy is to delimit the flow regimes considering the relevance of the viscous effects. This is done via the value of the Reynolds number, *Re*. This physical parameter compares the convective forces to the viscous forces of a fluid, $Re = \rho v L/\mu$ (where $\rho$ is the gas density, $v$ is the flow speed and $\mu$ is the gas dynamic viscosity). It will be seen later, in the Methodology section (Eq. (2)), that $Kn_r$, as defined using a frame fixed on the meteoroid, is a function of the *Re* number, and thus, using this scale the actual conditions for each event are more explicitly considered. Tsien (1946) noted the importance of these viscous effects and outlined a flow regime classification based on the comparison of the mean free path of the gas molecules ($l$) to the thickness of the boundary layer ($\delta$). This scale is then described as in Tsien (1946):

i) Free molecular regime, $Kn>10$;

ii) Transitional regime, $Re^{-1/2}<Kn<10$;

iii) Slip-flow regime, $10^{-2}\cdot Re^{-1/2}<Kn< Re^{-1/2}$;

iv) Continuum-flow regime, $Kn< 10^{-2}\cdot Re^{-1/2}$.

While the flow regime boundaries are fixed in the classical scale according to the intermolecular collision rate, the Tsien's scale accommodates for each event taking into account the viscous effect evolution. For instance, if *Re* increases, the transition and slip regime ranges shift to higher *Kn* numbers for that meteoroid. Conversely, as the *Re* decreases, the transitional and slip flow regime boundaries tend to shift to lower *Kn* values (and the continuum regime appears later). Note that these scales refer to the more general *Kn* definition (the reference frame moves with the gas flow), and the particulars derived from the use of another frame should be studied individually. In this study, in line with Bronshten (1983), the consideration of $Kn_r$ instead of *Kn*, which accounts for the mean free path of the reflected (evaporated) molecules relative to the impinging molecules ($l_r$) instead of the mean free path of the gas molecules (*l*), allows for the use of the two flow regimes scales (classical and Tsien's) described above. Additionally, Tsien (1946) originally suggested the classical scale to be used when the *Kn* is defined with the thickness of the boundary layer (Bronshten 1965).

Another classification was introduced by ReVelle (1993). He developed a meteoroid flight regime scale using *Kn* and three related parameters: a variation of the shape coefficient (effective mass/area), a variation of the ablation coefficient, and the height at which the kinetic energy has been reduced down to 1% of its initial entry value. This classification describes six different regimes. However, these parameters cannot be retrieved accurately from observations and thus the reliability of the results depends on the accuracy of the input data. This flight regime classification will not be accounted for in this study.

**1.2 The formation of the vapour cloud and the shock wave**

As the surrounding gas density increases, the number of impinging high energy particles becomes larger. The first layer of evaporated particles provides the meteoroid surface with a surrounding vapour cloud that screens the meteoroid from further high energetic impacts (also known as "hydrodynamic shielding"). The vapour cloud increases the number of the collisions, while the impinging particles are decelerated (Rajchl 1969, Bronshten 1983). When the mean free path of the vapour particles becomes an order of magnitude smaller than the meteoroid radius, the screening acts more efficiently (Popova et al. 2000). Besides, due to the reduction of high energy impacts, the atoms and ions within the hydrodynamic shielding cap can no longer be considered to be embedded in a hypersonic flow (see Bronshten 1965, 1983), and the hypersonic flight scenario becomes complex. Note that the simulations performed by Popova et al. (2000) for centimetre-size meteoroids show that the main dependences of the hydrodynamic shielding parameters are the size and the altitude of the meteoroid.

The vapour cloud virtually increases the cross-sectional area of the meteoroid (that collides with the atmosphere) by up to 2 orders of magnitude (Popova et al. 2000; Boyd 2000). When the vapour cloud reaches a pressure that exceeds that of the surrounding atmospheric gas (the vapour cloud is highly compressed), the vapour cloud expands like a hydrodynamic fluid into the surrounding, less dense environment (Popova et al. 2000). The outer layers of the cloud expand at supersonic speeds and a detached shock wave forms ahead of the body. The extent of the shock layer (defined as the space between the shock wave and the meteoroid surface) determines the amount of ionization and dissociation of the gas molecules (Bronshten 1965; Rajchl 1969). There is an extensive mathematical formulation and discussion on the physical phenomena that take place in the shock wave front, shock wave layer and meteor trail in Bronshten (1965). Along with this, a detailed scheme and a complete description of the meteor generated shock waves, the flow fields and the near wake can be found in Silber et al. (2017; 2018b).

According to the computational approach of Popova et al. (2000) and Boyd (2000), though based on several simplifying assumptions, the vapour cloud should appear during the transitional regime. This agrees with Rajchl (1969), who suggests that the vapour cloud should persist up until the beginning of the slip-flow regime. Nevertheless, identifying the moment when the meteor generated shock wave sets on is not fully understood. However, a more detailed discussion on this is beyond the scope of this paper, and the reader is referred to the comprehensive review on the topic of meteor generated shock waves (Silber et al. 2018b).

### 1.3 Linking the classical theory to observations

Observations of the meteor generated shock waves are complicated, and previous attempts using photometric measurements provided only preliminary conclusions (Rajchl 1972). While optical observations can be used to visually detect a meteor, this approach cannot provide solid evidence of the presence of the shock wave, especially for sub-centimetre and centimetre-sized meteoroids at high altitudes (e.g., the mesosphere-lower thermosphere or MLT region of the atmosphere). The high luminosity of the meteor phenomena, coupled with the fact that the shock front is very thin and attenuates very rapidly (Silber et al., 2017; 2018b), do not allow for direct optical detections of the shock wave (e.g., Schlieren photography). A quite different approach consists of surveying infrasound produced by the meteor generated shock waves.

Infrasound is low frequency (< 20 Hz) sound lying below the human hearing range and above the natural oscillation frequency of the atmosphere. Due to its very low attenuation rate, infrasound is an excellent tool for monitoring and studying impulsive sources in the atmosphere (e.g., ReVelle 1974; Silber et al. 2015; Silber & Brown 2019 and references therein). A shock wave, initially in the highly non-linear strong shock regime, eventually decays to a weakly nonlinear acoustic wave which could, given favourable conditions, be detected infrasonically at the ground (Silber et al. 2015). A theoretical approach to derive meteoroid parameters from infrasonic signatures, conceived by ReVelle (1974, 1976), was recently improved and subsequently validated (Silber et al. 2015) using a database of well constrained centimetre-sized meteoroids (Silber & Brown 2014). Using optical measurements and infrasound detections of bright meteors, Silber & Brown (2014) constrained the altitude of the meteor generated shock wave by finding the point along the meteor trajectory from which infrasound signal originated. Although this altitude is not diagnostic of the initial onset of the shock wave, it represents the earliest detected point at which the shock wave is proved to exist, which is an important pre-requisite for the purpose of our study. While there is strong evidence suggesting that in some cases the onset of meteor shock waves could take place much earlier than predicted by classical methodologies (Silber et al. 2017 and references therein), the Knudsen scale has never been verified against observations of centimetre-sized meteoroids.

In this study, we analyse the homogeneous database of 24 centimetre-sized meteoroids detected simultaneously by optical and infrasound systems, and published by Silber et al. (2015). However, constraining the meteoroid size (radii) could be challenging, as it may vary according to the methodology used (see, e.g. Gritsevich 2008c). Since the identification of the meteoroid flow regimes depends on this parameter, masses derived through five different approaches are accounted for in this study. First, an empirical law described by Jacchia et al. (1967) is used. It relates the following parameters to the meteoroid mass: the meteor magnitude in the photographic bandpass, the zenith angle of the radiant, and the speed at that point. Second, the photometric mass derivation method is applied as described in Ceplecha et al. (1998). It is known, that some portion of the kinetic energy lost by a meteoroid is converted to light emission, which can be mathematically expressed with the use of the luminous efficiency factor. The approach of Ceplecha et al. (1998) considers equation describing change in kinetic energy along with the assumption that a variation in the meteoroid velocity due to deceleration can be neglected compared to the loss of meteoroid mass. The magnitude of luminosity emitted by the meteor is then a function of the mass loss exclusively. Along with this, the rate of mass loss is assumed to be constant during ablation. The third photometric approach applied in the present work uses a more complex correlation between the fragmentation model and the light curve, described in Ceplecha & ReVelle (2005). A detailed description of the implementation particulars of these methods can be found in Silber et al. (2015), and thus will not be further described here. These three mass estimates will be hereafter referred to as JVB, IE (integrated energy) and FM, respectively, as previously defined and published in Table S3 of Silber et al. (2015). For comparative purposes, we also include the meteoroid mass estimates derived from the infrasound analyses (Silber et al. 2015) as the final two approaches. The fourth mass estimate is calculated from the observed information of the infrasonic signal period in the linear regime, and the fifth mass from the observed infrasonic signal period in the weak shock (ws) regime (ReVelle 1974, 1976). This will be described in Section 2.2, and further details can be found in Silber et al. (2015).

### 1.4 Implications of the identification of meteor flow regimes

Besides the simulations carried out by Boyd (2000) and Popova et al. (2000), the flow regimes of small meteoroids impacting the Earth at hypersonic velocities have not been studied in depth. These two studies tackled the problem from a numerical simulation approach. Campbell-Brown & Koschny (2004) developed a meteoroid ablation model for faint meteors under the free-flow regime conditions, and illustrated the differences in the meteoroid flow regimes with sizes up to one meter depending on whether the vapour cloud is taken into consideration or not. However, no study has described and validated the meteoroid flow regimes by means of observations that account for the existence of the hydrodynamic shielding.

As follows from Popova et al. (2000) and Silber et al. (2017), overdense meteors (as described in Silber et al. 2017, particles sized between $4 \cdot 10^{-3}$ m and a few centimetres) may reach the continuum flow regime below 90-95 km altitude as the flow pressure at that point will be smaller than the vapour gas pressure. It is well defined, though, that most meteoroids do ablate (which involves the possible onset of the vapour cloud and the shock wave) between 70 and 120 km; this region corresponds to the MLT region of the atmosphere. At these heights, the atmospheric conditions are dominated by large amplitude thermal and gravitational tidal waves which increase inner momentum of the fluid. Among other effects, this causes a rapid change in the gas molecular density which ultimately leads to a variation in the molecular mean free path.

Based on infrasound data analysis, it is possible to determine the earliest confirmed height along the meteor trail at which the shock wave is present. This knowledge can be used to determine the surrounding atmospheric gas conditions and ultimately the meteoroid flight flow regime. Moreover, since the shock wave is an indicator of the energy released by the event, the association of meteor flow regimes with the presence of a shock wave will provide relevant clues on the meteoroid flight parameters required to deposit energy in the upper atmosphere.

To our knowledge, the meteoroid data set of Silber et al. (2015) is the only well-documented and well-constrained set of centimetre-size events to-date. In this study, we aim to elucidate the complexities associated with the meteor flow regimes of bright meteors. Using the classical theory along with this homogeneous, observational data set of well-constrained meteoroid events recorded both optically and infrasonically, we aim to determine and validate the flow regimes of centimetre-sized meteoroids in the upper atmosphere. In order to get a deeper insight on the suitability of this approach, both the classical and the Tsien (1946) Knudsen scales are implemented to determine the flow regimes. We also examine whether these two *Kn* scales can be employed as useful proxies in determining the flow regimes of meteoroids in the cm-size range in future studies. This also allows us to elucidate the flow regimes associated with an apparent early onset of meteor generated shock waves by linking the observations to a theoretical approach. To our knowledge, none of these points have been addressed before.

The paper will continue with a description of the infrasound methodology and the *Kn* calculation in Section 2. The results and discussion are summarized in Section 3. Finally, the conclusions of this work are presented in Section 4.

## 2. METHODOLOGY

### 2.1 The data set – background

Our data set is taken from Silber et al. (2015). While the detailed methodology outlining data collection, reduction and analyses pertaining to the data set was published in Silber and Brown (2014), we briefly summarize important points here for clarity. The meteors in the data set were recorded simultaneously by all-sky cameras (the All-Sky and Guided Automatic and Realtime Detection (ASGARD) network) and infrasound array (the Elginfield Infrasound Array (ELFO)), which are the part of the regional fireball observations network located in Southwestern Ontario (Canada).

The advantages of having both optical and infrasound systems within the same network, and thus close together, are twofold. First, given favourable conditions, some meteors (such as those analysed in this study), can be recorded by both optical and infrasound systems simultaneously. Second, it is more likely to detect direct arrivals, or infrasound sources within ~300 km of the receiver. The relevance of this lies in the fact that there is a rapid decrease of the infrasound signal-to-noise ratio (SNR) for events that originate too far from the infrasound array (> 300 km). Provided that the shock wave typically forms at high altitudes (Popova et al. 2000; Silber et al. 2017), the atmospheric conditions along the propagation path can adversely affect the signal and therefore hinder the detection efficiency of infrasound. Thus, direct arrivals are less likely to suffer from irreversible changes (Silber & Brown 2014; 2019). Only about 1% of optically detected centimetre sized meteoroids are also captured by infrasound (Silber & Brown 2014).

Our data set consists of only the best constrained events for which at least one infrasound source height is accurately obtained, have reliable optical measurements, and do not show abrupt deceleration or fragmentation. Several cases for which two infrasound sources are obtained are also included in this study, but only the earliest source is considered. This is because only the highest altitude associated with the shock wave is relevant to the analysis of the flow regimes, as this is where the most uncertainty exists. Low altitudes (e.g. below 70 km) are usually associated with the continuum flow, where the verification is then no longer a practical task.

## 2.2 Derivation of meteoroid sizes from masses

The estimation of the meteoroid characteristic size, its radius ($r$), is not straightforward. This value is derived from the meteoroid masses. The masses used in this study have been derived using the five different methods, as described in the Introduction, three of them based on the analysis of the photometric light curve produced by the meteor and the remaining two using infrasound techniques. The infrasound masses are calculated using Eq. (8) in Silber et al. (2015):

$$M_{infra} = (\pi \rho_m / 6)(R_0 / Ma)^3 \qquad (1)$$

where $\rho_m$ is the meteoroid density and $R_0$ the blast radius. The blast radius is proportional to the product of the meteoroid diameter ($d$) and the Mach number ($R_0 \simeq d \cdot Ma$), and it is defined as the distance between the shock source to the point where the overpressure (the excess pressure over the local atmospheric pressure generated by the shock wave) approaches the local atmospheric pressure. Thus, it is a way of determining the instantaneous energy deposition. Kinetic energy and $R_0$ are interconnected (Figure 1a), especially if there is no abrupt deceleration or gross fragmentation that would skew the magnitude of $R_0$ (see Silber et al. 2015 for further discussion). Indeed, as shown in Figure 1a, none of the events analysed here undergo fragmentation or abrupt deceleration, which attests to the suitability of the data set for the purpose of our study. The blast radius can be obtained through correlating the observed infrasonic signal period with the modelled period in the linear and weak shock regimes (for a more detailed discussion, see Silber et al. 2015). It should be stated that, while infrasound is a reliable tool for detecting meteors and estimating the source function, it has not been validated sufficiently well for the purpose of estimating the meteoroid masses. Hence, infrasound masses are often either under- or overestimated compared to photometric masses. Despite this shortcoming, we include meteoroid radius estimates from infrasonic masses for the purpose of direct comparison and for the sake of completeness.

One source of uncertainty to be considered when calculating the five radius estimates (photometric and infrasonic) is that meteoroids do not have a fixed bulk density value. While this value is usually assumed to be fairly similar to a certain reference density according to the meteorite classification, other parameters such as the micro- and macro-porosity or case specific mineral inclusions can alter it significantly (Britt & Consolmagno 2003; Babadhaznov & Kokhirova 2009; Meier et al. 2017).

Possible meteoroid associations to well-studied annual meteor showers were explored by Silber & Brown (2014). Previous studies of known meteor showers could provide additional clues on the meteoroid density. However, since only five of the events in our data set show such a relationship, providing insufficient statistics, for this work the possible density values for each meteor shower are disregarded. From the observational data, Silber et al. (2015) retrieved the PE parameter (see table S4 in Silber et al. 2015) described in Ceplecha & McCrosky (1976). The use of this parameter as a meteor classification criterion has been widely adopted (e.g. Brown et al. 2013a). The range of densities assigned to each PE value relies on the statistics built up with the density derivation for each meteoroid using a dynamic analysis of the trajectory of accurately observed meteors; however, individual density errors may ultimately affect the statistics of the result. The PE values for some meteors of the current data set lead to the meteoroid density values of 270 kg/m$^3$. Such a value is smaller than that of water ice (916.8 kg/m$^3$). Though these density values might be possibly depending on the packing factor of fractal-like structures (see e.g., Blum et al. 2006), typical meteoroid bulk densities are usually larger (e.g., common chondritic meteorite bulk density ranges between 3000 – 3700 kg/m$^3$; see Consolmagno & Britt 1998; Flynn et al.1999; Wilkison & Robinson 2000). On the other hand, as per the classical classification of meteoroids accepted for stony bodies, a reasonable bulk density approximation corresponds to the value of 3500 kg/m$^3$ (Levin 1956). This value has been widely in use (see, e.g., Halliday et al. 1996; Ceplecha 1998; Gritsevich 2008b; Gritsevich 2009; Bouquet et al. 2014) and it is thus chosen for this work. Note that this value could be large for fragile meteoroids as discussed in Britt & Consolmagno (2003), who suggest density close to 2500 kg/m$^3$ for carbonaceous chondrites. Nonetheless, the assumption of either value does not significantly affect the resulting $Kn_r$ number. The meteoroid data set under this study consists of cm-sized bodies whose exact characteristic size may show only slight variation, according to the mass and density chosen. Furthermore, this variation could be neglected, as the Knudsen number is principally affected by the characteristics (velocity, density and temperature) of the incoming flow. In the scenario studied in this work, the high energy collisions with the ambient species are effective in slowing down the ablated species in the meteor flow field. This consequently leads to high ranges of temperature and density in the shock layer, which play the main role in varying the value of $Kn_r$. Thus, the most critical input parameter in this analysis is the incoming gas flow velocity.

The characteristic meteoroid radii were derived for each of the five mass estimates by considering a spherically shaped object of the same mass and bulk density. It is evident that the mass estimates obtained from each methodology (photometric and infrasound) differ notably due to intrinsic assumptions associated with each. We will discuss shortly what the implications are to the overall results in this study (see the Section 3). The radii, along with other parameters obtained from the meteor infrasound detection and luminous path observations by Silber et al. (2015) are shown in Table 1. Note that all the five meteoroid sizes vary from r ~ 0.18 to r ~8.8 cm. The spread in meteoroid radii as a function of altitude is shown in Figure 1b.

### 2.3 Calculation of the Knudsen number

We now turn our attention to the approach to obtain the flow regimes from classical considerations, as applicable to the data set at hand. As already stated in the Introduction, the meteoroid reaches a point at which the surrounding screening vapour gas expands like a hydrodynamic fluid into the surrounding, less dense environment (Popova et al. 2000). This causes the atmospheric gas density to adapt abruptly to the expanding vapour gas. This creates a shock wave through which the atmospheric gas increases its pressure and temperature. The equations of Rankine-Hugoniot relate this change between the gas state at both sides of the detached shock wave. These equations can be applied if one-dimensional compressible, inviscid and adiabatic fluid is assumed. Thus, they do not consider viscosity effects, radiation or conduction heat transfer, nor gravitational acceleration.

Using these relations, the gas conditions behind the detached (if the Mach number of the gas flow behind the shock layer is subsonic) shock wave can be retrieved. It is important to note that the density and temperature jump of the shock wave strongly depends on the adopted $\gamma$ value. Thus, increasing or decreasing $\gamma$ could vary the magnitude of this jump. While the best approach would be to vary $\gamma$ according to the atmospheric conditions and the physical scenario, the dynamical changes in the value of $\gamma$ in the flow field can only be tracked through sophisticated numerical simulations. Even so, the existing numerical models are unable to accurately describe the hypervelocity flow conditions associated with meteoroids propagating at velocities greater than about 35 km/s, especially in the upper atmosphere, where the object might be on the boundary of the transitional flow. Thus, in our study, the gas is assumed to be calorically ideal, with the constant ratio of specific heat ($\gamma = c_p/c_v$) equal to 1.4 (this is the value for an ideal diatomic gas). This assumption is generally considered to be a valid approximation for explosive sources with a narrow channel (when the shock wave can be approximated as a cylindrical line source, see Taylor 1950) including meteoroid entry problems, and as such is also employed in other studies (e.g., Popova et al. 2000; Zhdan et al. 2007; Sansom et al. 2015; Chen et al. 2017). The reasoning for such approach is that the rarefied ambient density (e.g., the MLT) decreases the value of $\gamma$, while presence of strong radiative phenomena (associated with meteors) increases the value of $\gamma$. While this might be an oversimplification, any other assumptions implemented in the analytical approach and the classical theory could introduce additional uncertainties and skew the results.

The atmospheric conditions, density and temperature, of the incoming gas flow are estimated using an empirical atmospheric model. For this study, the NRLMSISE-00 atmospheric model (Picone et al. 2002) was chosen. This model provides the atmospheric profile above a specific geocentric location (longitude, latitude and ground altitude) for a required date and time and is among recommended for the use in meteor analysis (Lyytinen & Gritsevich 2016). We use the geographical location of the infrasound array and the infrasound wave arrival time for each event (Table 1) in order to retrieve the atmospheric conditions from the NRLMSISE-00 model. These are then used as the input parameters in the Rankine-Hugoniot equations to obtain the flow conditions in the shock layer and eventually allow the derivation of the *Ma*, *Re* and *Kn* numbers.

The meteor events in our data set have shock source height uncertainties that range between 0.3 km and 4.2 km (see column 3 in Table 2), although for most of the cases this uncertainty is $\leq 1$ km. For such a limited height uncertainty, the surrounding atmospheric gas conditions will not show large variations and therefore it is possible to assume that the gas pressure, density and temperature values are fixed.

Once the atmospheric conditions of the incoming gas flow are determined (temperature, density and velocity), the sound speed and the Mach number upstream and downstream relative to the shock wave, and the gas state in the shock layer are calculated. Note that a normal front shock wave has been assumed. In principle, the bow shock wave tends to wrap around the meteoroid; however, the Mach cone angle, defined as the angle between the body movement direction and the normal vector of the shock wave, is equal to the arcsin(1/*Ma*), and thus it deviates only marginally from zero for the incoming gas flow.

The resulting atmospheric gas conditions behind the shock wave are used to derive the Knudsen number. As discussed in the Introduction, the $Kn_r$ is the most suitable Knudsen number description for meteor physics problems. Equation (2) shows the relationship between the $Kn_r$ and the gas physical variables (Bronshten 1983):

$$Kn_r = \frac{1}{Re} \cdot \frac{\overline{V_e}}{c_s} = \frac{1}{Re} \cdot \frac{1}{c_s} \cdot \left(\frac{8T_w R}{\pi M}\right)^{\frac{1}{2}} = \frac{1}{Re} \cdot \frac{(8T_W)^{\frac{1}{2}}}{\sqrt{\gamma \pi T}} = \frac{\mu}{v_\infty \rho r} \cdot \left(\frac{8T_w}{\gamma \pi T}\right)^{\frac{1}{2}} \qquad (2).$$

Here, $c_s$ is the local speed of sound, $\overline{V_e}$ is the average velocity of the vaporizing molecules (Bronshten 1965), $R$ is the universal constant of the gases, $M$ is the molar mass of the gas, $T_w$ is the meteoroid's surface temperature, $\gamma$ is the constant ratio of specific heat, $\mu$ is the gas dynamic viscosity, $v_\infty$ is the velocity of the incoming gas flow, $\rho$ is the gas density, $r$ is the equivalent radius of the meteoroid (derived assuming a spherical body), and $T$ is the gas temperature. Note that according to Eq. (2) $Kn_r$ can be expressed in terms of the $Re$ number and the local speed of sound.

The derivation of the $Kn_r$ (Eq. (2)) involves the previous knowledge of a set of variables. The density and the temperature of the incoming gas are calculated behind the shock wave. The gas flow conditions upstream and downstream of the shock wave can be found in Table 2 (note that the upstream and downstream, respectively, refer to the flow regions ahead and behind a reference point, which in this case is the shock wave).

The dynamic viscosity is a function of the gas temperature, and it is given by Sutherland (1893):

$$\mu = \frac{1.458 \cdot 10^{-6} \sqrt{T}}{1 + \frac{110.4}{T}} \qquad (3).$$

The velocity of the incoming gas flow is the velocity of the meteoroid when the frame of reference is set on the meteoroid surface. For simplicity, this velocity was assumed to be equal to the initial velocity observed along the meteor luminous trajectory path. While this value will remain temporally constant only for those fast meteors within the study data set that experience little deceleration, it will be argued later that the $Kn_r$ results are not largely affected and this assumption is valid. Additionally, meteoroids typically undergo notable deceleration at lower altitudes, where the atmospheric density is greater. Thus, at altitudes investigated here, deceleration can be assumed to be negligible. Furthermore, as stated in Silber et al. (2015), the meteoroids in our data set did not undergo abrupt deceleration, as that was one of the pre-requisites of the weak shock model validation.

Finally, there is no unique methodology to determine the meteoroid surface temperature. Indeed, it is a challenging issue. It is generally assumed that upon the onset of the ablation, the main evaporation phase begins once the temperature reaches 2500 K (Ceplecha et al. 1998; Boyd 2000; Popova et al. 2001; Jenniskens 2006), and it shall not largely increase afterwards as the kinetic energy is mainly employed in the ablation process itself. On the other hand, using emission spectroscopy techniques, Borovička (1993, 1994) and Trigo-Rodríguez et al. (2003, 2004) compared synthetic spectra with the observed meteor spectra and found an excellent match for most lines. They determined that there were two separate range of temperatures that could match the two differentiated spectral components that the meteors produced at 3500 - 5000 K for most of the excited composition elements, and at around 10 000 K for some specific ionized elements. As the infrasound analysis reveals the altitude at which the shock wave originated (but not the earliest point at which the meteoroid started generating the shock wave upon entering the atmosphere), a conservative approach was used assuming that the meteoroid surface temperature is close to 2500 K. Furthermore, as the shock source altitude was constrained by Silber et al. (2015) to within ± 1 km for more than half of the cases (although eleven events have the altitude uncertainty of up to 4.2 km, see Table 2, column 3), there exists a difficulty in accurately determining the level of evolution of the ablation process of the meteoroid. It should be noted, though, that the temperature rise in the shock layer will reach and even exceed ~ $10^6$ K. Hence, depending on material properties and velocity of the meteoroid, the meteoroid surface temperature $T_w$ will be two or three orders of magnitude smaller than the gas flow temperature, and as stated by Eq. (2), variations between $T_w$ ~2500 - 5000 K will not largely affect the rate $T_w/T$. The remaining uncertainty is well within the uncertainties in the radii size.

## 3. RESULTS AND DISCUSSION
### 3.1 Analysis of the Knudsen number results

The results of the $Kn_r$, $Re$ and flow field calculations are summarized in Table 3. We show the relations between $Kn_r$ and various quantities; these are altitude (Figure 2a), kinetic energy (Figure 2b), meteoroid velocity (Figure 2c) and meteoroid mass (Figure 2d). For clarity, $Kn_r$ values derived from all five mass estimates (JVB, IE, FM, linear period and shock wave period) are plotted. Note that Figure 2 offers an insight into how these variables behave at the different flow regimes of the classic scale. For instance, no meteoroid is observed in the transitional flow regime ($10^{-1} < Kn_r < 10$) when the infrasound masses are considered. The linear relationship between the shock source and the $Kn_r$ shown in Figure 2a demonstrates that for well constrained cm-sized meteoroids, the formation of the hydrodynamic shielding may affect the meteoroid flow regime by shifting it to lower $Kn_r$. Figure 2 also provides a visual demonstration of how errors in the mass or size calculations affect the meteoroid flow regime. As expected, if the meteoroid velocity is kept constant, but the mass (and consequently the effective radius) is increased, the flow regime shifts to lower Knudsen numbers for the shock source altitudes observed.

The amount of kinetic energy released at the shock source height shows little variation when all the masses and their respective $Kn_r$ are compared. Figure 2b indicates a slight shift toward higher $Kn_r$ of those meteoroids with lower energies. However, care must be given here, as the statistically small meteoroid data set might lead to a weak relationship. It can, however, be acknowledged that the energy deposition at the shock altitudes (50 to 100 km) varies by three orders of magnitude, from $10^3$ kJ to $10^6$ kJ. The combination of different values of the velocity and entry angle affects how the meteoroid releases energy and produces infrasound that can be detected on the ground (Silber & Brown 2014). The results obtained here expand this discussion and allow us to determine the flow regime associated with the point along the meteor trajectory at which the energy was deposited (and subsequently recorded by infrasound). The results (Table 3) suggest that the shock waves could, in principle, form prior to the continuum flow regime and mainly during the slip-flow regime (or even the transitional if the classical scale is considered). We attribute this to the formation of the hydrodynamic shielding, which, as explained in the Section 1.2, acts to increase the effective size of the meteor cross-section (Bronshten 1983; Popova et al. 2000; Campbell-Brown & Koschny 2004; Silber et al. 2018b). While this result suggests that infrasound can be used to obtain relevant meteoroid flight parameters, more sophisticated numerical models (yet to be developed) are recommended to further investigate our assertion and to determine the earliest possible point at which the shock wave forms when a meteoroid undergoes strong ablation in rarefied flow conditions.

The results shown in Figure 2c show that the shock wave associated with the fastest meteoroids is detected when these bodies are between the transitional and slip flow regimes according to the classical scale. We will see later that if the Tsien's scale is used (Table 3), all meteoroids are within either the slip or continuum flow regime. Note that for these fast meteoroids, the shock wave is detected at higher altitudes than usually expected for a typical meteoroid (see Table 2). Our results corroborate the results of Popova et al. (2000) which suggest that in fast moving meteoroids, the flow regime will be shifted upwards and the shock wave should, indeed, form at higher altitudes. Moreover, the presence of the vapour cap in strongly ablating meteoroids will also affect the flow regime (Popova et al. 2000). This might explain why, typically, fast meteoroids can be visually observed sooner than slow meteoroids. Conversely, slow meteoroids will reach lower altitudes before the shock wave can be detected (see, e.g. Silber et al. 2018a).

Figure 2d illustrates that infrasound masses have a tendency to towards lower $Kn_r$, while photometric masses show a spread across all $Kn_r$ and thus exhibit a weak relationship. In principle, this tendency is due to the already mentioned mass overestimation through infrasound analyses. A plausible explanation for this apparent discrepancy is the formation of hydrodynamic shielding, which could, in principle, affect the energy deposition and thus the size of the blast radius. In fact, Eq. (1) assumes that no or very little ablation is taking place, which, in reality, is rarely the case. Therefore, the infrasound mass derived from the energy deposition (and the blast radius) might not necessarily correspond to the physical mass of the object itself. In some cases, both infrasonic and photometric JVB masses may differ notably relative to the photometric IE and FM masses. In principle, the larger the meteoroid cross section, the larger the number of collisions against atmospheric particles, and the sooner the vapour cap is formed. Consequently, larger masses (which represent larger sizes if the same value of density is assumed) are consistent with lower $Kn_r$, which agrees with the results shown in Figure 2d. Finally, the broad distribution of IE and FM masses is expected, as the meteoroid mass (or size) is only one of several factors (e.g., altitude, velocity) controlling $Kn_r$.

Another important point to note is, as discussed by Popova et al. (2000), that the vapour cap will shift the meteoroid continuum flow regime to higher altitudes. This is because the presence of the vapour cap effectively increases the cross section of the region colliding with air molecules.

**3.2 Validation of the results with two Knudsen classification scales**

Matching the resulting $Kn_r$ to a specific level of the classical Knudsen scale is somewhat subjective. The uncertainties in the mass (and thus size) derivation lead to different values. As shown in Table 3, despite minor differences, the three $Kn_r$ numbers obtained from the JVB, IE and FM photometric masses show little variation in terms of the flow regimes. The task of assigning a flow regime when $Kn_r$ value lies near the flow regime boundaries is strictly related to the precision at which we accept these boundaries to be sharp, although, in reality, this transition is not necessarily sharp. Slight $Kn_r$ variations around these 'edges' are merely nominal and so if two different masses lead to the same flow regime, this is accepted as the current state. According to this scheme, 33% of the meteoroid data set is in the transitional regime, 46% in the slip-flow and the remaining 21% has already reached the continuum flow. Note that these statistics are only used to get a preliminary view of the phenomenology; indeed, for some events the $Kn_r$ is on the boundary between the slip and continuum regimes. A similar discussion can be applied to the Tsien (1946) scale. In this case, the meteoroid data set shows the following distribution: 88% in the slip-flow regime and 12% in the continuum flow regime.

In view of these results, the use of three different masses (JVB, IE and FM) for each meteoroid proves that the effect of the assumed meteoroid bulk density value is not critical. Even in the case of the largest difference between mass estimates (meteoroid ID 20110808), the $Kn_r$ number does not vary by much (this is so in both scales). Furthermore, the effect of the extreme meteoroid bulk densities (according to the PE scale: 270 and 7000 kg/m$^3$) were explored showing that for the lowest density case (270 kg/m$^3$), the flow regime may vary for 33% of the events in the classical scale and 12% in the Tsien's scale. In the classical scale, these events shift either from the transitional to the slip flow regime, or from the slip flow to the continuum regime. However, it should be mentioned that most of these cases were previously lying in-between the two flow regimes using the assumed stony meteoroid bulk density. Moreover, the use of the Tsien's scale shows that only three cases move to the continuum regime, but once again, these were close to the boundary cases. The use of the highest bulk density (7000 kg/m$^3$) leads to the variation in two cases in the classical scale and one case in Tsien's scale, all shifting from the continuum to the slip flow regime. These small variations due to the bulk density are expected as the effect of either the mass or the bulk density only affects the meteoroid characteristic size which was determined to be in well constrained.

Even though the meteoroid data set in this study is not considered to undergo abrupt deceleration (Silber et al. 2015), we examine a certain level of deceleration to overcome the effect of any measurement inaccuracy in our results. This is because the meteoroids, by their very nature will undergo ablation (more or less strong), which in turn will result in deceleration, especially at lower altitudes. A new value of this velocity was applied assuming a deceleration of 30% (this value exceeds typical deceleration values for centimeter sized meteoroids, see Jenniskens et al. (2011), but will help in understanding the effect of the velocity on the derivation of the $Kn_r$). It must be emphasized that the entry velocity used here was that obtained at the first luminous observed point of the meteor trajectory; at that point, the shock wave may have already been formed. Although the shock source heights shown in Table 2, column 2, indicate points within the luminous trajectory, these points represent the earliest point in the trajectory at which the shock wave was detected. However, the shock wave could certainly have appeared even earlier.

According to this, our results show that there are only two different event flow regimes that change in the classical scale and the Tsien's scale. Thus, introducing deceleration in order to account for any inaccuracies in the calculation of the entry velocity does not affect our results, and only two events shift from the continuum flow to the slip flow regime. The reason behind this apparent flow regime invariability is the energy conversion at the shock front. The transformation of the kinetic energy of the incoming gas flow at the shock front elevates both the temperature and the density in the shock layer. However, on one hand, the gas density, which remains too low, and the small size of the body still balance the increase due to the velocity variation (see Eq. (2)); on the other hand, these high temperature conditions provide dynamic viscosity values that are well below 1. Consequently, the $Re$ number does not vary significantly. However, this small variation still alters the boundaries of the Tsien's scale (see the comments in the Introduction section) which tend to shift towards higher $Kn_r$. Using this new velocity value, all meteoroids in our data set propagate under the slip-flow conditions, except for one case, which remains in the continuum regime. Although this new velocity, accounting for deceleration, is more extreme than should occur in the MLT, we use it to test the parameter space bounds in our calculations.

The two $Kn_r$ numbers derived from the infrasound linear and weak shock wave period masses are quite similar (see columns 5 and 6 in Table 3), and generally different from the JVB, IE and FM $Kn_r$ numbers. We reiterate that the JVB masses do remarkably differ from the IE and FM masses and, in several cases resemble the mass of the infrasound linear and weak shock methodologies. This could open the discussion on whether the JVB methodology is accurate enough. A previous study that critically compared photometric masses to those derived through dynamic approach (Gritsevich 2008a), also demonstrated that more work is required to reconcile the apparent differences. However, its use helps understanding the effects of possible erroneous measurements on the $Kn_r$ determination. The use of exclusively the infrasound masses leads to 54% of the events in the slip-flow regime and a 46 % in the continuum regime according to the classical scale. As for the Tsien's scale, 79% of the cases are in the slip-flow and the remaining 21% in the continuum regime. Despite the small size of the data set, it can be recognized that these results agree with those derived using the classical scale. In fact, except for one case, all the five masses provide the same flow regime when the Tsien's scale is in use. This is because, as derived from the previous discussion and Eq. (2), the value of $Kn_r$ is strongly influenced by the entry velocity and the atmospheric gas conditions at the height where the shock wave is detected. These parameters are principally gathered in the $Re$ number. Moreover, the importance of the viscous effects that are already relevant in the expanding vapour gas is held in the $Re$ number; this suggests that the use of the Tsien's scale is more appropriate in this study. Conversely, the use of the classical scale does not take into account the actual physical scenario that viscosity may create. It is therefore interesting to note that there could be other more complex combinations of fluid dynamics dimensionless characteristic parameters that could delimit more appropriately the meteoroid flight regimes.

The results provided indicate that the flight flow regime for most of the meteoroids in this data set is between the lower half of the slip-flow regime and the beginning of the continuum regime (the Tsien's scale is assumed here). If it could be further verified that the shock wave forms in these regimes, it would be in agreement with the work of Rajchl (1972). However, there is no clear evidence of that and the suggestion of Probstein (1961), by which the shock wave may gradually form once past half of the transitional regime, cannot be rejected. Future studies should be done in this regard.

We note, while the assumption that $\gamma = 1.4$ might be a simplification, it still provides reasonable results that are consistent with the observations. For example, as expected, no meteor event is found to be in the free molecular flow at altitudes that suggest the presence of the shock wave. The consideration of varying $\gamma$ is best suited for numerical models, although some modeling studies did apply $\gamma = 1.4$ and found that the main dependences of the vapour (hydrodynamic shielding) parameters, and consequently the temperature and density jumps, are the size and the altitude of the meteoroid (see Popova et al. 2000 and Section 1.2). Also, the consideration of an ablating centimeter-sized meteoroid entering at velocities up to 73 km/s is very different and profoundly more complex than, for example, a much larger re-entry vehicle at significantly lower velocities (e.g., 7 km/s) (see Silber et al. 2018b for discussion).

Finally, the current study uses a reference frame located on the surface of the meteoroid (see the discussion in the Introduction), thus moving with the body (i.e. local phenomena). However, although well beyond the scope of this paper, it could also be possible to combine this information ($Kn_r$, local) with the information that arises from the global picture, that is, the $Kn$ study of the immersed body (meteoroid plus the vapour gas cap) in the surrounding gas flow. The global and local outcome retrieved from studying both parameters could be of interest in analyzing individual cases and should be considered in future studies.

**3.3 Implications of the shock wave information in the study of the flow regimes**

Infrasound observations shed light on only a portion of the whole event. As stated by Silber & Brown (2014), infrasound indicates the earliest confirmed point at which the shock wave originated, but the question what the maximum altitude is at which the shock waves can form remains open. This is indeed a source of uncertainty, but it also validates the fact that meteor shocks form at much higher altitude than they would by theoretically considering their size alone. For instance, it can be found within the meteoroid data set that some members show high altitude infrasound, which is in line with previous studies for centimeter-sized bodies (Brown et al. 2007; Silber & Brown 2014 and references therein). Thus, there is already a shock wave at these altitudes. Even in those cases, this study shows that the Tsien's scale appropriately describes the flow regimes even for these high-altitude events. Note that thus far, no observational or modeling studies have resolved the intricacies associated with the formation of a shock wave in the MLT region for meteoroids traveling at hypervelocity and in the rarefied flow conditions. Furthermore, at present, there are no numerical models that account for all meteor associated phenomena (e.g. ablation, radiation) in the rarefied flow conditions. Thus, this should be the focus of future studies.

Popova et al. (2000) discussed the flow regimes for a Leonid meteoroid with entry velocity of ~72 km/s. As stated before, the meteoroid propagates under the free-molecular flow conditions until the onset of intense evaporation at lower heights. Due to this mass loss, the vapour cloud (or hydrodynamic shielding) forms gradually, and when the mean free path within the vapour cloud is much smaller than the meteoroid radius ($l_v \sim 0.1r$) the screening acts more efficiently and the meteoroid is no longer in the free -molecular regime. The vapour cap is then formed and the meteoroid enters the transitional flow regime between the free flow and the continuum regimes. Note, however, that Popova et al. (2000) use the classical scale and so $l_v \sim 0.1r$ represents the "boundary" between the transitional and slip-flow regimes when $Kn_r$ is considered. Note also that the transition regime mentioned by Popova et al. (2000) should really account for the slip-flow regime in the classical scale as it is derived from the use of the classical scale ($0.01 < l_r/r < 0.1$). One additional consideration to be noted, as stated by Popova et al. (2000), is that once the vapour temperature exceeds 4000 K, the cloud becomes optically thick, and so it hinders the release of the increasing energy within the vapour cloud in the form of radiation. This latter effect may increase the vapour pressure as described in Section 1.2 of this work, leading to the formation of a shock wave.

This study deals with the meteoroid flow regimes from an observational aspect and upon the formation of the shock wave. We use an adaptation of Figure 1 of Popova et al. (2000) to plot our meteoroid data set and to put our results in perspective. This is shown in Figure 3. This figure includes the boundaries and flow regimes as described by Popova et al. (2000) for $10^{-2}$ to 10 cm sized Leonid meteoroids considering a dense vapour cloud in front of the body. The altitude used to plot our data is that at which the shock wave is detected (the shock source height), whereas as for the meteoroid size, a mean value for the estimated sizes (see Table 3) through various methodologies is chosen. Figures 3a,c show the average meteoroid radius for the JVB, IE and FM masses, while Figures 3b,d display the mean value for the infrasound linear and weak shock wave period derived sizes. Note that, as the altitude is a fixed value, the position of the meteoroids in each panel of Figure 3 may only vary along the abscissa according to the methodology used in the meteoroid radius derivation. The intense evaporation line, the beginning of the vapour cloud formation ($l_v \sim 0.1r$), the limit below which the vapour temperatures ($T_v$) exceeds 4000 K, and the boundary for the continuum flow regime for the Leonid meteoroid studied in Popova et al. (2000) are also plotted.

In order to provide a deeper insight in the results we have used different shapes and colors in Figure 3 to indicate the flow regime of each meteoroid as derived in our study (Table 3), namely: blue circles illustrate that the meteoroid is in the transitional regime, orange triangles represent the slip-flow regime and green squares represent the continuum regime. Since two *Kn* scales are under analysis, we have plotted in the panels on the left (Figure 3a, b) our meteoroid flow regime results as derived from the use of the classical scale, while the panels on the right (Figure 3c, d) illustrate the meteoroid regimes when the Tsien's scale is considered. Note again that the flow regimes areas labelled in the plots are those obtained by Popova et al. (2000) for their Leonid **meteoroids**, and so they do not represent the calculated flow regimes for our meteoroid data set.

The first thing to be noted is that the presence of a shock wave indicates that our meteoroids are located below the line of $T_v \sim 4000$ K, which is indeed the case. However, the existence of a shock wave changes the conditions in the vapour cloud and thus the meteoroid could reach the lower *Kn* earlier. Although the division of flow regimes by Popova et al. (2000) does not directly apply to our data set, it can serve as the basis for visualization. It can be seen that the continuum flow regime is not reached by nearly any of the meteoroids in our data set. Indeed, the slip-flow regime is achieved at a wider range of heights. It should be noted that the delimitation of the flow regimes by Popova et al. (2000) applies to Leonid meteoroids with a roughly fixed entry velocity of 72 km/s, whereas our meteoroid data set shows a range of entry velocities (13.5 – 71.2 km/s, as shown in Table 2). However, our data set contains three meteoroids with entry velocities

close to 72 km/s, namely: 20060805, 20070125 and 20081107. Two of these, 20060805 and 20070125, show a $Kn_r$ that is on the boundary between the transitional and slip-flow regimes (classic scale) and as such, their position on Figure 3a is closer to the free-molecular flow outlined by Popova et al. (2000). These two meteoroids are located around 20 km below the free-molecular flow delimitation line, thus supporting the statement that the appearance of shock wave is suggestive of the alteration of classically defined meteor flow regimes.

The use of the mean meteoroid size, though not completely accurate, is still representative of the realistic scenario. Using either end member estimate of the radius for a given meteoroid would move the position of the data point along the x-axis (Figure 3) to the right or the left (note that the x-axis is in logarithmic scale). The x-axis error bars in Figure 3 indicate the standard error from the mean values. As stated before, the masses derived from the infrasound linear and weak shock periods are very similar, and thus this error is small. However, the meteoroid sizes derived using the JVB methodology show larger discrepancies when compared to the IE and FM results; this causes the large error bars. If the JVB masses were disregarded in the study, the meteoroid radii in the figure would be practically fixed. Nevertheless, as the sizes of the meteoroids in the data set are well constrained and the flow regimes are determined, these large error bars are useful to indicate the extent of uncertainty that might be expected in these types of studies. It can be stated from Figure 3 that the results derived from the infrasound linear and weak shock period radii are generally within the size errors of the mean photometric radius (JVB, IE and FM).

The formation of the hydrodynamic shielding and eventually the shock wave alters the mean free path in the vicinity of the meteoroid, and therefore the flow regime conditions. This implies a dynamical scenario that could be difficult to track using a fixed classification of the classical Knudsen scale. As per our results, we suggest that the formation of the vapour cap (or hydrodynamic shielding) should be re-evaluated in the definition of the meteoroid flow regimes. In fact, the vapour cap plays an important role in the generation of the shock wave, and the extent of this role should be the scope of more sophisticated models (yet to be developed) and future studies. In these terms, the introduction of a classification scheme that accounts for changes in the surrounding conditions, such as Tsien's scale, seems more reliable.

## 4. CONCLUSIONS

This study has explored the utility of meteoroid infrasound to unravel new clues on the atmospheric flight regime of centimeter-sized bodies. Coupled with optical observations, infrasound provides conclusive evidence of the existence of meteor generated shock wave at a given altitude. As the meteoroid penetrates deeper into the atmospheric layers, the incoming flux of atmospheric particles increases, and the ablation process starts. Sporadic gas molecular collisions become more regular, triggering an intense vaporization process. This leads to the formation of a vapour cloud in front of the meteoroid. Once the pressure of this cloud exceeds that of the surrounding atmospheric gas, it expands, and a detached shock wave is formed in front of the meteoroid. The acoustic by-product of the shock wave (infrasound) can be detected under certain conditions from ground-based instruments. The use of that information has been implemented here to reach the following conclusions:

i) Previous works based on infrasound analysis demonstrated that the infrasound study could positively identify the earliest point at which it can be claimed that a shock wave is present. Furthermore, those studies also suggest that the meteor shock wave could form much earlier than predicted by classical methodologies. On the other hand, despite the limited information provided, infrasound seems to be a robust means to determine the flow regime of meteoroids. This study provides the first observational verification of the Knudsen scale using information obtained through infrasound for a data set of centimeter-sized meteoroids. This data set represents the only well-documented and well-constrained set of such events to-date.

ii) Our results are consistent with the use of a reference frame attached to the meteoroid body, in contrast to the gas flow attached reference frame. Such approach is not only more convenient, but also more representative of realistic conditions. Moreover, it has been shown that the flow regimes could be considered within boundaries delimited as function of several fluid dynamic dimensionless parameters (i.e. *Kn*, *Re*, *Ma*). The results reinforce the theoretical approach that claims that a scale based in the *Kn* and *Re* numbers illustrates the physics of the problem more accurately. The differences between the flow regimes derived from the theoretical and observational approaches have been discussed. While no strong conclusion could be derived as the formation height of the shock wave cannot be determined yet, this study suggests that the shock wave for cm-sized meteoroids is already formed in the slip-flow regime (or even late transitional flow regime).

iii) This study also explored whether the use of information derived from different meteoroid observation techniques could lead to similar results. In this sense, photometric measurements provide the robust means of estimating cm-sized meteoroid masses (under condition of negligible deceleration). While infrasound alone does not provide sufficient insight into meteoroid masses, it remains an excellent tool in monitoring and detection of meteors. Moreover, infrasound measurements, when coupled with other techniques, provide useful estimates in meteor flow regimes, and thus could serve as another mode of validation. This study shows that simultaneous observations of meteors, using both infrasound and photometric techniques can provide relevant clues on the meteoroid flight regimes and the energy deposition at the point of origin of shock wave.

iv) Our study confirms that the formation of a vapour cap shifts the flow regimes upward and acknowledges the necessity of developing new and more sophisticated models to describe the flow regimes of meteoroids encountering the Earth's atmosphere. These new models should also constrain and evaluate the impact on the hydrodynamic shielding in those events where a strong ablation takes place. This fact would eventually play a relevant role on the formation of the meteor generated shock wave and shift the flow regimes. Several questions remain open and shall be the scope of future research: once the maximum height at which the shock wave can form is more accurately determined, would the flow regime vary by much; what is the most suitable flow regime scale; and is there any use in combining the information obtained using different reference frames ($Kn$ vs $Kn_r$). A natural step towards further refinement would include numerical studies and determination how the dynamic changes in the hypervelocity flow field might affect the flow regimes.

## 5. ACKNOWLEDGMENTS

MMI and JMTR thank the support of the Spanish grant AYA2015-67175-P. EAS gratefully acknowledges the Natural Sciences and Engineering Research Council of Canada (NSERC) Postdoctoral Fellowship program for supporting this project. MG acknowledges support from the ERC Advanced Grant No. 320773, and the Russian Foundation for Basic Research, project nos. 16-05-00004, 16-07-01072, and 18-08-00074. Research at the Ural Federal University is supported by the Act 211 of the Government of the Russian Federation, agreement No 02.A03.21.0006. The authors acknowledge being a part of the network supported by the COST Action TD1403 "Big Data Era in Sky and Earth Observation". This study was done in the frame of a PhD. on Physics at the Autonomous University of Barcelona (UAB) under the direction of Dr. Maria Gritsevich and Dr. Josep M. Trigo-Rodríguez. The authors thank the anonymous referee for the valuable comments that helped improve this paper.

| Date | Hour | Minute | Seconds | H begin [km] | H end [km] | Mass (JVB) [g] | Mass (IE) [g] | Mass (FM) [g] | Mass Infrasound (linear p.) [g] | Mass Infrasound (weak shock p.)[g] | Radius (JVB) [cm] | Radius (IE) [cm] | Radius (FM) [cm] | Infra Radius (linear p.) [cm] | Infra Radius (weak shock p.)[cm] |
|---|---|---|---|---|---|---|---|---|---|---|---|---|---|---|---|
| 20060419 | 7 | 5 | 56 | 72.0 | 47.7 | 107.4 | 23.5 | 20.0 | 94.9 | 75.9 | 1.94 | 1.17 | 1.11 | 1.86 | 1.73 |
| 20060805 | 8 | 38 | 50 | 126.4 | 74.5 | 5927.6 | 432.9 | 74.0 | 2292.7 | 1038.3 | 7.39 | 3.09 | 1.72 | 5.39 | 4.14 |
| 20061104 | 3 | 29 | 29 | 89.9 | 65.8 | 459.9 | 12.5 | 12.0 | 1.6 | 1.1 | 3.15 | 0.95 | 0.94 | 0.48 | 0.42 |
| 20070125 | 10 | 2 | 5 | 119.2 | 88.5 | 9.5 | 2.7 | 0.9 | 2924.5 | 1375.2 | 0.86 | 0.57 | 0.39 | 5.84 | 4.54 |
| 20070727 | 4 | 51 | 58 | 96.2 | 70.6 | 2583.9 | 91.5 | 63.0 | 816.4 | 428.6 | 5.61 | 1.84 | 1.63 | 3.82 | 3.08 |
| 20071021 | 10 | 26 | 25 | 130.8 | 81.7 | 57.5 | 10.6 | 4.3 | 2005.9 | 967.5 | 1.58 | 0.90 | 0.66 | 5.15 | 4.04 |
| 20080325 | 0 | 42 | 3 | 76.2 | 32.8 | 2912.0 | 792.9 | 917.0 | 133.0 | 105.4 | 5.83 | 3.78 | 3.97 | 2.09 | 1.93 |
| 20080511 | 4 | 22 | 17 | 95.2 | 77.3 | 85.8 | 5.2 | 8.0 | 1603.0 | 822.5 | 1.80 | 0.71 | 0.82 | 4.78 | 3.83 |
| 20080812 | 8 | 19 | 29 | 105.7 | 82.0 | 0.2 | 0.1 | 0.1 | 125.0 | 70.6 | 0.22 | 0.18 | 0.20 | 2.04 | 1.69 |
| 20081028 | 3 | 17 | 35 | 81.2 | 41.1 | 309.8 | 79.6 | 110.0 | 56.7 | 46.8 | 2.76 | 1.76 | 1.96 | 1.57 | 1.47 |
| 20081102 | 6 | 13 | 26 | 96.5 | 62.6 | 663.9 | 53.3 | 18.0 | 112.1 | 69.5 | 3.56 | 1.54 | 1.07 | 1.97 | 1.68 |
| 20081107 | 7 | 34 | 16 | 113.5 | 81.5 | 0.4 | 0.2 | 0.1 | 332.7 | 208.7 | 0.30 | 0.22 | 0.20 | 2.83 | 2.42 |
| 20090428 | 4 | 43 | 37 | 83.5 | 38.0 | 3086.5 | 784.1 | 330.0 | 686.0 | 489.3 | 5.95 | 3.77 | 2.82 | 3.60 | 3.22 |
| 20090523 | 7 | 7 | 25 | 95.9 | 72.4 | 2.7 | 0.7 | 2.2 | 125.0 | 81.1 | 0.57 | 0.36 | 0.53 | 2.04 | 1.77 |
| 20090812 | 7 | 55 | 58 | 108.5 | 80.4 | 20.6 | 3.4 | 1.8 | 41.8 | 25.1 | 1.12 | 0.61 | 0.50 | 1.42 | 1.20 |
| 20090917 | 1 | 20 | 38 | 85.7 | 72.4 | 20.7 | 6.6 | 8.5 | 112.7 | 71.8 | 1.12 | 0.77 | 0.83 | 1.97 | 1.70 |
| 20100421 | 4 | 49 | 43 | 108.5 | 74.6 | 861.5 | 45.7 | 17.0 | 534.3 | 314.6 | 3.89 | 1.46 | 1.05 | 3.32 | 2.78 |
| 20100429 | 5 | 21 | 35 | 105.7 | 89.9 | 0.9 | 0.2 | 0.3 | 283.7 | 159.8 | 0.40 | 0.25 | 0.26 | 2.68 | 2.22 |
| 20100530 | 7 | 0 | 31 | 96.0 | 78.3 | 1.2 | 0.3 | 0.3 | 1281.4 | 682.6 | 0.43 | 0.27 | 0.26 | 4.44 | 3.60 |
| 20110520 | 6 | 2 | 9 | 95.7 | 84.1 | 21.3 | 2.3 | 2.5 | 555.6 | 304.7 | 1.13 | 0.54 | 0.55 | 3.36 | 2.75 |
| 20110630 | 3 | 39 | 38 | 100.5 | 71.7 | 527.5 | 18.0 | 10.0 | 15.6 | 9.3 | 3.30 | 1.07 | 0.88 | 1.02 | 0.86 |
| 20110808 | 5 | 22 | 6 | 86.6 | 39.9 | 9990.9 | 2586.4 | 1003.0 | 1465.3 | 1045.3 | 8.80 | 5.61 | 4.09 | 4.64 | 4.15 |
| 20111005 | 5 | 8 | 53 | 96.2 | 64.5 | 6.8 | 2.6 | 20.0 | 17.7 | 12.2 | 0.77 | 0.56 | 1.11 | 1.06 | 0.94 |
| 20111202 | 0 | 31 | 4 | 97.0 | 53.8 | 18.0 | 8.8 | 9.0 | 1413.9 | 1075.8 | 1.07 | 0.84 | 0.85 | 4.59 | 4.19 |

**Table 1.** Basic data retrieved from the meteor infrasound detection and luminous path observations. Photometric meteoroid masses taken from Silber et al. (2015) are calculated as described in Jacchia et al. (1967), JVB; using the kinetic energy as in Ceplecha et al. (1998), IE; using the fragmentation model and the light curve described in Ceplecha & ReVelle (2005), FM. Infrasonic masses (linear period and weak shock period) has been calculated using Eq. (2) and following the work of Silber & Brown (2014). The meteoroid radii are derived from these masses. The columns are organized as follows: (1) meteoroid ID (which coincides with the date of its detection); (2-4) the time at which the infrasonic wavetrain reached the detector; (5-6) the beginning and ending heights of the meteor luminous path; (7-11) the meteoroid masses derived using five different methodologies; (12-16) the results of the meteoroid radius calculation (using the masses listed in previous columns. Except for the infrasound masses and meteoroid radii, all the other data shown in this table was previously published by Silber et al. (2015).

| ID | Shock Source Height [km] | Error S.S. Height [km] | Flow Conditions Upstream | | | | | Flow Conditions Downstream | | | | | Atmospheric Viscosity [kg/(m·s)] |
|---|---|---|---|---|---|---|---|---|---|---|---|---|---|
| | | | $V_{entry}$ [km/s] | T [K] | Density [g/cm3] | Sound Speed [m/s] | Mach | T [K] | Density [g/cm3] | Mach | Sound Speed [m/s] | V [m/s] | |
| 20060419 | 54.4 | 1.1 | 14.2 | 255.1 | 6.461E-07 | 320.0 | 44.32 | 97651.9 | 3.867E-06 | 0.3785 | 6260.5 | 2369.4 | 0.0005 |
| 20060805 | 101.4 | 0.4 | 67.5 | 191.8 | 3.379E-10 | 277.5 | 243.32 | 2208143.7 | 2.026E-09 | 0.3780 | 29770.3 | 11252.6 | 0.0022 |
| 20061104 | 77 | 1.1 | 30.3 | 218.5 | 2.461E-08 | 296.1 | 102.18 | 443808.2 | 1.477E-07 | 0.3781 | 13346.5 | 5045.7 | 0.0010 |
| 20070125 | 102.7 | 0.5 | 71.2 | 181 | 3.396E-10 | 269.5 | 264.31 | 2458858.3 | 2.037E-09 | 0.3780 | 31415.0 | 11874.2 | 0.0023 |
| 20070727 | 85 | 1.5 | 26.3 | 165.6 | 8.244E-09 | 257.8 | 102.05 | 335505.5 | 4.946E-08 | 0.3781 | 11604.3 | 4387.1 | 0.0008 |
| 20071021 | 101.2 | 1.4 | 64.3 | 185.6 | 4.722E-10 | 272.9 | 235.59 | 2003159.0 | 2.832E-09 | 0.3780 | 28354.9 | 10717.6 | 0.0021 |
| 20080325 | 61.6 | 0.6 | 13.5 | 237.2 | 2.414E-07 | 308.6 | 43.75 | 88516.3 | 1.445E-06 | 0.3785 | 5960.5 | 2255.9 | 0.0004 |
| 20080511 | 94.6 | 0.4 | 23.5 | 188.5 | 1.418E-09 | 275.1 | 85.58 | 268631.1 | 8.502E-09 | 0.3781 | 10383.6 | 3926.0 | 0.0008 |
| 20080812 | 87.9 | 0.8 | 56.6 | 174.5 | 4.952E-09 | 264.6 | 213.87 | 1552152.8 | 2.970E-08 | 0.3780 | 24959.6 | 9434.4 | 0.0018 |
| 20081028 | 52.7 | 3.6 | 15.4 | 252.1 | 6.79E-07 | 318.1 | 48.41 | 115132.0 | 4.063E-06 | 0.3784 | 6797.8 | 2572.1 | 0.0005 |
| 20081102 | 85 | 0.5 | 30.1 | 209.7 | 7.222E-09 | 290.1 | 103.75 | 439121.2 | 4.329E-08 | 0.3781 | 13275.8 | 5019.0 | 0.0010 |
| 20081107 | 81.9 | 0.6 | 71.6 | 214.4 | 1.137E-08 | 293.3 | 244.08 | 2483801.8 | 6.821E-08 | 0.3780 | 31573.9 | 11934.3 | 0.0023 |
| 20090428 | 70.9 | 1.1 | 21.2 | 217.7 | 7.448E-08 | 295.6 | 71.72 | 217940.1 | 4.466E-07 | 0.3782 | 9352.7 | 3536.8 | 0.0007 |
| 20090523 | 78.1 | 2.3 | 29.9 | 194.4 | 2.772E-08 | 279.3 | 107.04 | 433293.2 | 1.661E-07 | 0.3780 | 13187.5 | 4985.5 | 0.0010 |
| 20090812 | 80.6 | 0.3 | 58.7 | 186.4 | 1.78E-08 | 273.5 | 214.61 | 1669465.6 | 1.068E-07 | 0.3780 | 25885.6 | 9784.4 | 0.0019 |
| 20090917 | 76.6 | 2.1 | 24.2 | 206.5 | 3.051E-08 | 287.9 | 84.06 | 283912.6 | 1.830E-07 | 0.3781 | 10674.9 | 4036.2 | 0.0008 |
| 20100421 | 86.3 | 0.8 | 45.9 | 190.7 | 6.602E-09 | 276.7 | 165.91 | 1020839.5 | 3.961E-08 | 0.3780 | 20241.8 | 7651.4 | 0.0015 |
| 20100429 | 93 | 1.9 | 47.7 | 186.3 | 2.019E-09 | 273.4 | 174.44 | 1102456.7 | 1.210E-08 | 0.3780 | 21035.4 | 7951.3 | 0.0015 |
| 20100530 | 92.7 | 2.4 | 29.3 | 171.7 | 1.973E-09 | 262.5 | 111.61 | 416063.9 | 1.183E-08 | 0.3780 | 12922.6 | 4885.3 | 0.0009 |
| 20110520 | 94.5 | 0.7 | 22.5 | 183.6 | 1.465E-09 | 271.5 | 82.89 | 245429.9 | 8.786E-09 | 0.3781 | 9925.1 | 3752.7 | 0.0007 |
| 20110630 | 87.7 | 0.5 | 29.8 | 161.4 | 5.042E-09 | 254.5 | 117.08 | 430369.9 | 3.025E-08 | 0.3780 | 13142.9 | 4968.5 | 0.0010 |
| 20110808 | 63.6 | 0.3 | 25.5 | 230.9 | 2.229E-07 | 304.4 | 83.76 | 315236.4 | 1.336E-06 | 0.3781 | 11248.3 | 4253.0 | 0.0008 |
| 20111005 | 77.8 | 4.2 | 28.5 | 208.7 | 2.276E-08 | 289.4 | 98.47 | 393697.5 | 1.365E-07 | 0.3781 | 12570.5 | 4752.4 | 0.0009 |
| 20111202 | 64 | 0.6 | 27.6 | 234.6 | 1.474E-07 | 306.9 | 89.94 | 369261.7 | 8.842E-07 | 0.3781 | 12174.1 | 4602.8 | 0.0009 |

**Table 2.** Shock wave analysis: shock source height and its error values derived from infrasound study; and gas flow conditions upstream and downstream calculated using the Rankine-Hugoniot equations. Columns are organized as follows: (1) the meteoroid ID; (2-3) the source height of the shock wave and the associated error; (4) the entry velocities (which are used to estimate the incoming gas flow velocity, as described in the main text); (5-8) the gas temperature, gas density, sound speed and Mach number upstream, respectively; (9-14) the downstream conditions in the following order: (9) gas temperature, (10) gas density, (11) Mach number, (12) sound speed, (13) gas velocity and (14) the gas dynamic viscosity.

| Date | Kn_r (JBV) | Kn_r (IE) | Kn_r (FM) | Kn_r (linear p.) | Kn_r (weak shock p.) | Re (JBV) | Re (IE) | Re (FM) | Re (linear p.) | Re (weak shock p.) | Classical Scale | Tsien Scale | Classical Scale for infrasound masses | Tsien Scale for infrasound masses |
|---|---|---|---|---|---|---|---|---|---|---|---|---|---|---|
| 20060419 | 0.000 | 0.001 | 0.001 | 0.000 | 0.001 | 391.0 | 235.6 | 223.3 | 375.2 | 348.3 | Continuum | Slip | Continuum | Continuum |
| 20060805 | 0.049 | 0.117 | 0.210 | 0.067 | 0.087 | 0.8 | 0.3 | 0.2 | 0.6 | 0.4 | Transitional | Slip | Slip | Slip |
| 20061104 | 0.004 | 0.012 | 0.012 | 0.023 | 0.026 | 24.2 | 7.3 | 7.2 | 3.7 | 3.2 | Slip | Slip | Slip | Slip |
| 20070125 | 0.393 | 0.599 | 0.862 | 0.058 | 0.075 | 0.1 | 0.1 | 0.0 | 0.6 | 0.5 | Transitional | Slip | Slip | Slip |
| 20070727 | 0.007 | 0.021 | 0.023 | 0.010 | 0.012 | 14.4 | 4.7 | 4.2 | 9.8 | 7.9 | Slip | Slip | Slip | Slip |
| 20071021 | 0.172 | 0.302 | 0.408 | 0.053 | 0.067 | 0.2 | 0.1 | 0.1 | 0.8 | 0.6 | Transitional | Slip | Slip | Slip |
| 20080325 | 0.000 | 0.001 | 0.001 | 0.001 | 0.001 | 438.9 | 284.5 | 298.6 | 156.9 | 145.2 | Continuum | Slip | Continuum | Slip |
| 20080511 | 0.137 | 0.348 | 0.301 | 0.052 | 0.064 | 0.8 | 0.3 | 0.4 | 2.1 | 1.7 | Transitional | Slip | Slip | Slip |
| 20080812 | 0.134 | 0.162 | 0.146 | 0.014 | 0.017 | 0.3 | 0.3 | 0.3 | 3.2 | 2.6 | Transitional | Slip | Slip | Slip |
| 20081028 | 0.000 | 0.000 | 0.000 | 0.001 | 0.001 | 584.9 | 371.9 | 414.2 | 332.0 | 311.6 | Continuum | Continuum | Continuum | Continuum |
| 20081102 | 0.011 | 0.025 | 0.035 | 0.019 | 0.023 | 8.0 | 3.5 | 2.4 | 4.4 | 3.8 | Slip | Slip | Slip | Slip |
| 20081107 | 0.034 | 0.045 | 0.052 | 0.004 | 0.004 | 1.0 | 0.8 | 0.7 | 10.0 | 8.6 | Slip | Slip | Continuum | Continuum |
| 20090428 | 0.001 | 0.001 | 0.002 | 0.001 | 0.002 | 138.1 | 87.4 | 65.5 | 83.6 | 74.7 | Slip | Slip | Continuum | Slip |
| 20090523 | 0.018 | 0.027 | 0.019 | 0.005 | 0.006 | 4.9 | 3.1 | 4.6 | 17.6 | 15.3 | Slip | Slip | Continuum | Slip |
| 20090812 | 0.007 | 0.013 | 0.016 | 0.006 | 0.007 | 6.2 | 3.4 | 2.8 | 7.9 | 6.6 | Slip | Slip | Continuum | Slip |
| 20090917 | 0.010 | 0.015 | 0.013 | 0.006 | 0.007 | 10.7 | 7.3 | 7.9 | 18.8 | 16.1 | Slip | Slip | Continuum | Slip |
| 20100421 | 0.007 | 0.019 | 0.026 | 0.008 | 0.010 | 8.0 | 3.0 | 2.2 | 6.8 | 5.7 | Slip | Slip | Continuum | Slip |
| 20100429 | 0.216 | 0.339 | 0.332 | 0.032 | 0.039 | 0.2 | 0.2 | 0.2 | 1.7 | 1.4 | Transitional | Slip | Slip | Slip |
| 20100530 | 0.332 | 0.519 | 0.553 | 0.032 | 0.040 | 0.3 | 0.2 | 0.2 | 2.7 | 2.2 | Transitional | Slip | Slip | Slip |
| 20110520 | 0.220 | 0.463 | 0.450 | 0.074 | 0.091 | 0.5 | 0.2 | 0.3 | 1.5 | 1.3 | Transitional | Slip | Slip | Slip |
| 20110630 | 0.017 | 0.051 | 0.062 | 0.054 | 0.064 | 5.2 | 1.7 | 1.4 | 1.6 | 1.3 | Slip | Slip | Slip | Slip |
| 20110808 | 0.000 | 0.000 | 0.000 | 0.000 | 0.000 | 611.2 | 389.6 | 284.1 | 322.3 | 288.0 | Continuum | Continuum | Continuum | Continuum |
| 20111005 | 0.016 | 0.023 | 0.011 | 0.012 | 0.013 | 5.5 | 4.0 | 7.9 | 7.5 | 6.7 | Slip | Slip | Slip | Slip |
| 20111202 | 0.002 | 0.002 | 0.002 | 0.000 | 0.000 | 49.1 | 38.7 | 39.0 | 210.6 | 192.3 | Continuum | Continuum | Continuum | Continuum |

**Table 3.** Knudsen numbers, Reynolds numbers and meteoroid flow regime analysis: (1) event ID; (2-6) $Kn_r$ as derived from the five possible masses discussed in Section 3; (7-11) the *Re* number using these five masses; the flow regime according to the classical scale (see the Introduction) and the scale described in Tsien (1946) as obtained from the JVB, IE and FM masses (12-13), and the masses derived from the infrasound detected signal (linear and weak shock period) (14-15).

**Figure 1**

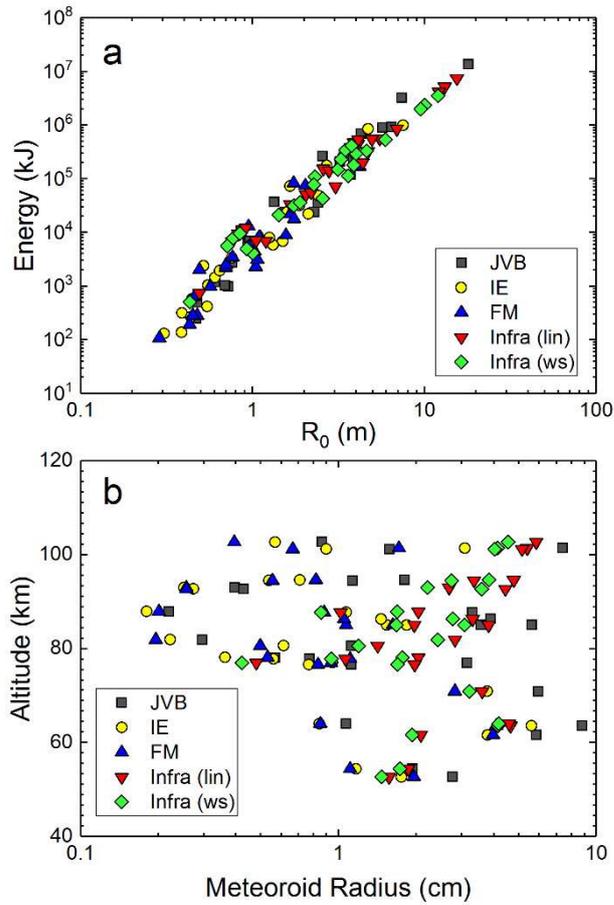

**Figure 1:** (a) The meteoroid kinetic energy plotted against infrasound blast radius ($R_0$) for the five masses derived in this study; (b) The shock source altitude plotted against meteoroid radii, as retrieved from the JVB, IE, FM, and infrasound masses (from linear and shock weak methodologies).

**Figure 2**

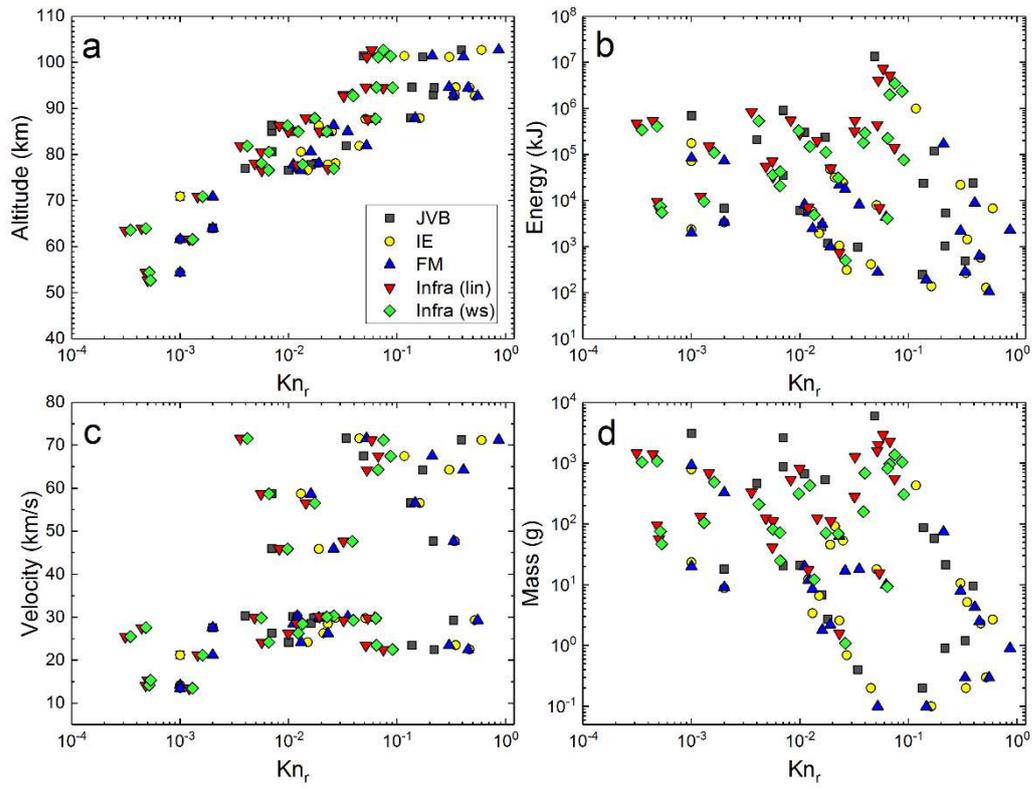

**Figure 2:** Relation between $Kn_r$, as derived from the five masses retrieved from observations (JVB, IE, FM, linear period and shock wave period) with: (a) the shock source altitude; (b) the kinetic energy; (c) the meteoroid entry velocity; (d) the meteoroid mass. Note that the legend in panel (a) is applicable to the rest of plots (b-d).

**Figure 3**

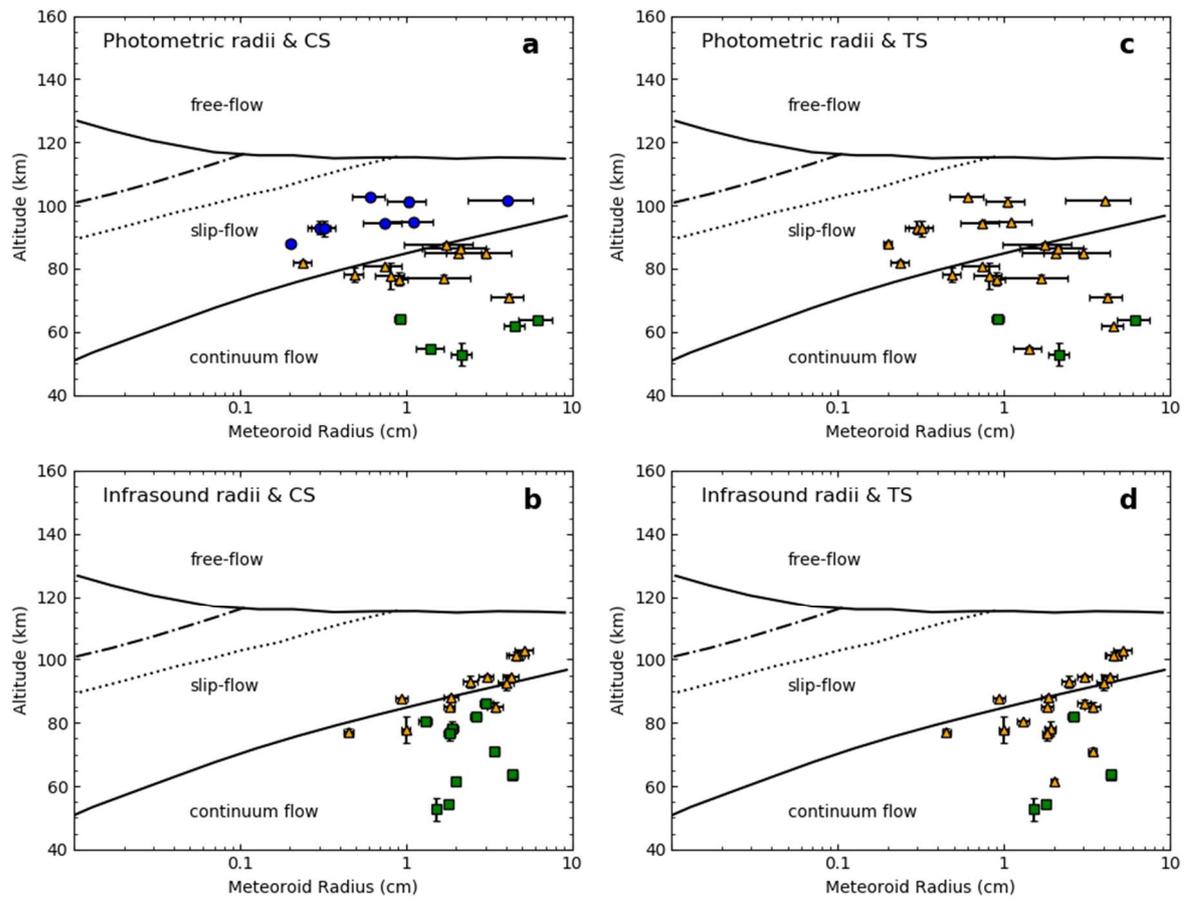

**Figure 3.** Adaptation of Figure 1 of Popova et al. (2000). The lines and regions are as in Popova et al. (2000): the intense evaporation line (continuous top line) and the continuum flow (continuous bottom line) boundary for the Leonids (0.01 cm sized meteoroids with entry velocities around 72 km/s); the boundary that indicates the moment the mean free path ($l_v$) becomes 0.1 times the meteoroid radius ($l_v \sim 0.1r$) or, conversely, the beginning of the slip-flow regime when the classical scale is in use (dashdotted line); and the line below which the vapour cloud temperature ($T_w$) exceeds 4000 K (dotted line). The flow regime regions for these Leonid meteoroids as derived by Popova et al. (2000) are labelled. The mean meteoroid radii from the JVB, IE and FM photometric masses are shown in panels (a) and (c). While panels (b) and (d) plot the results for the mean meteoroid radii derived from the infrasound methodologies (linear and weak shock periods). The flow regimes as derived from the two scales analysed in this study are represented by data points with distinct colours and shapes. Blue circles are used for meteoroids in the transitional flow regimes. Orange triangles represent those meteoroids in the slip-flow regime. Green squares indicate the continuum regime. The panels on the left (a, b) account for the flow regimes when the classical scale (CS) is considered, whereas in the panels on the right (c, d), the meteoroid flow regimes are based on the results using the Tsien's scale (TS).

Finally, the horizontal error bars represent the standard error from the mean, and the altitude error as described in Table 2. Note that some error bars are small and contained within the data points.